\documentclass{PoS}
\usepackage{lineno}
\usepackage{graphicx}
\title{Spectral analysis of Markarian 421 and Markarian 501 with HAWC}

\ShortTitle{Spectral analysis of Markarian 421 and Markarian 501 with HAWC}

\author{\speaker{Sara Couti\~no de Le\'on}, Alberto Carrami\~nana Alonso and Daniel Rosa-Gonz\'alez for the HAWC collaboration\thanks{Complete list of authors at http://www.hawc-observatory.org/collaboration/icrc2017.php}\\
        Instituto Nacional de Astrof\'isica, \'Optica y Electr\'onica, Puebla, Mexico.\\
        E-mail: \email{sara@inaoep.mx}, \email{alberto@inaoep.mx}, \email{danrosa@inaoep.mx}}

\abstract{The Hight Altitude Water Cherenkov (HAWC) Gamma-Ray Observatory monitors the gamma-ray sky in the energy range from 100 GeV to 100 TeV and has detected two very high energy (VHE) blazars: Markarian 421 (Mrk 421) and Markarian 501 (Mrk 501) in 1.5 years of observations. In this work, we present the spectral analysis above 1 TeV of both sources using a maximum likelihood method and an artificial neural network as an energy estimator. The main objectives are to constrain the spectral curvature of Mrk 421 and Mrk 501 at $\sim$5 TeV using the EBL models from Gilmore et al. (2012) and Franceschini et al. (2008).}

\FullConference{35th International Cosmic Ray Conference - ICRC217-\\
		10-20 July, 2017\\
		Bexco, Busan, Korea}

\begin{document}

\section{Introduction}

Blazars are a type of radio-loud active galactic nuclei (AGN) with their jets pointing towards the observer \cite{blazarjet}, they are divided in two categories: flat-spectrum radio quasars (FSRQ) and BL Lac objects. In this work we focus in two BL Lacs: Markarian 421 (Mrk 421) and Markarian 501 (Mrk 501). In general, BL Lacs are characterized by the lack of emission lines, that is why the majority of the BL Lacs have been discovered in the X-ray to $\gamma$-rays frequency range \cite{bllacs}. BL Lacs are characterized by a non-thermal continuum spectra with a broad low-frequency component in the radio-UV/X-ray frequency range and a high-frequency component from X-rays to $\gamma$-rays, and are often characterized by a variability through the electromagnetic spectrum. Mrk 421 and Mrk 501 are the nearest ($z=0.031$ and $z=0.034$ respectively) BL Lacs. They are classified as high-synchrotron peaked (HSP) blazars since their synchrotron peak is located at $E>40$ eV \cite{HBL}. Mrk 421 and Mrk 501 are being detected at $\sqrt{\mbox{TS}}=40.39$ and $\sqrt{\mbox{TS}}=25.96$, respectively  by the High Altitude Water Cherenkov (HAWC) Observatory which operates in the 100 GeV to 100 TeV energy range \cite{catalog}. In this work we present the preliminary spectral properties of these sources at very high energies (VHE) using $\sim$1.5 years of observations with HAWC and an energy estimator based on an artificial neural network.

\section{Data}\label{Data}

The HAWC Observatory is a ground-based TeV gamma-ray observatory in the state of Puebla, Mexico at an altitude of 4100 m a.s.l. The detector continuously measures the arrival time and direction of gamma-ray primaries within its 2 sr field of view. It is most sensitive to gamma-ray energies ranging from 100 GeV - 100 TeV. Cuts on the data can be applied to differentiate gamma-ray air showers from the large cosmic-ray background, also a reconstruction of the air shower in the detector is performed to obtain the direction of the primary $\gamma$-ray and its size. For more details about the detector performance read \cite{Crab,LC}.

The data used for this analysis goes from November 2014 to June 2016 comprising 543 transits. In order to divide the data into energy bins, an event-by-event energy-reconstruction algorithm has been developed using an artificial neural network (NN) to estimate the $\gamma$-rays primary energy. A NN is a function that maps quantities associated with an event (input variables) to some output variables, in this case $E$ which is the primary energy of the shower. The input variables are divided in three categories that are used to characterize the shower energy: the multiplicity in the detector that is a proxy of the energy reaching the ground, the containment of the shower that indicates by how much the multiplicity in the detector needs to be scaled up, and the atmospheric attenuation of the shower, which indicates how much energy was lost on the way to the ground. For more details see \cite{NN}. The energy bins are shown in the Table \ref{energybins}.

\begin{table}
\begin{center}
\begin{tabular}{ccc}
\hline 
Bin & Log(E/GeV)  & Energy Range\\ 
\hline 
a & 2.50-2.75 & 316-562 GeV \\ 
b & 2.75-3.00 & 562 GeV-1.00 TeV  \\ 
c & 3.00-3.25 & 1.00-1.77 TeV\\ 
d & 3.25-3.50 & 1.77-3.16 TeV\\ 
e & 3.50-3.75 & 3.16-5.62 TeV\\ 
f & 3.75-4.00 & 5.62-10.0 TeV\\ 
g & 4.00-4.25 & 10.0-17.78 TeV\\ 
h & 4.25-4.50 & 17.78-31.62 TeV\\ 
i & 4.50-4.75 & 31.62-56.2 TeV\\ 
i & 4.75-5.00 & 56.2-100 TeV\\ 
k & $>$ 5.00  & $>$ 100  TeV\\ 
\hline 
\end{tabular} 
\end{center}
\caption{\small{Energy bins. In this work, the bins a and k were not used due to the low statistic.}} \label{energybins}
\end{table}

\section{Methodology}\label{Methodology}

For the spectral analysis we used the HAWC maximum-likelihood framework (LiFF) described in \cite{LiFF} to fit the photon fluxes during the 543 transits. The fit includes the attenuation due to the extragalactic background light (EBL) in the spectral modeling. Given a intrinsic model spectrum, LiFF attenuates the predicted fluxes which are then convolved with the detector response function given by simulations, and then compares it to data. As described in \cite{LiFF}, the likelihood ratio test is the comparison between the likelihood of a source model and the likelihood for a background-only model (null hypothesis), and the test statistics is defined as the ratio between these quantities, which is maximized depending on the free parameters of the source model:
\begin{equation}
\mbox{TS}= 2\left[\ln\mathfrak{L}(\mbox{Source model})- \ln\mathfrak{L}(\mbox{Background-only model})\right].
\end{equation}
The $\sqrt{\mbox{TS}}$ values are interpreted as significances (sigmas) and as pointed in \cite{catalog}, we consider a source detected when TS$>25$ or $\sqrt{\mbox{TS}}>5$.

In order to fit the spectral shape of the sources, we first use all the energy bins leaving all the spectral parameters free to vary. It is important to mention that the uncertainties given here are statistical from the fit since the NN energy estimator is still under development. 

\section{Results}\label{Results}

\subsection{Mrk 421}
We used EBL models from \cite{Gil} and \cite{Fran} to perform the fit using two different spectral assumptions: a power law (PL) and a power law with an exponential energy cut-off (PL+CO). The result of the fits are shown in the Table \ref{mrk421_global}. For both EBL models and using all the energy bins, the PL+CO model was preferred with $\Delta\mbox{TS}=39.62$ using the EBL model from \cite{Gil} and $\Delta\mbox{TS}=25.09$ for the EBL model from \cite{Fran}.

\begin{table}
\begin{center}
\begin{tabular}{cccc}
\hline 
EBL model & $\Gamma$ & $E_c$ [TeV] & $\Delta\mbox{TS}$\\ 
\hline 
Gilmore et al. (2012)      & $2.02\pm0.09$ & $4.75\pm0.70$ & 39.62\\
Franceschini et al. (2008) & $2.04\pm0.07$ & $5.28\pm0.87$ & 25.09 \\
\hline 
\end{tabular} 
\end{center}
\caption{\small{Fit results for Mrk 421 using EBL models from Gilmore et al. 2012 \cite{Gil} and Franceschini et al. 2008 \cite{Fran}, $\Gamma$ is the spectral index, $E_c$ is the energy cut-off and $\Delta\mbox{TS}$ is the TS difference between the spectral models assumptions. For this source, the PL+CO spectral model is preferred independently from the EBL model.}}\label{mrk421_global}
\end{table}

\begin{figure}
\centering
\includegraphics[scale=0.5]{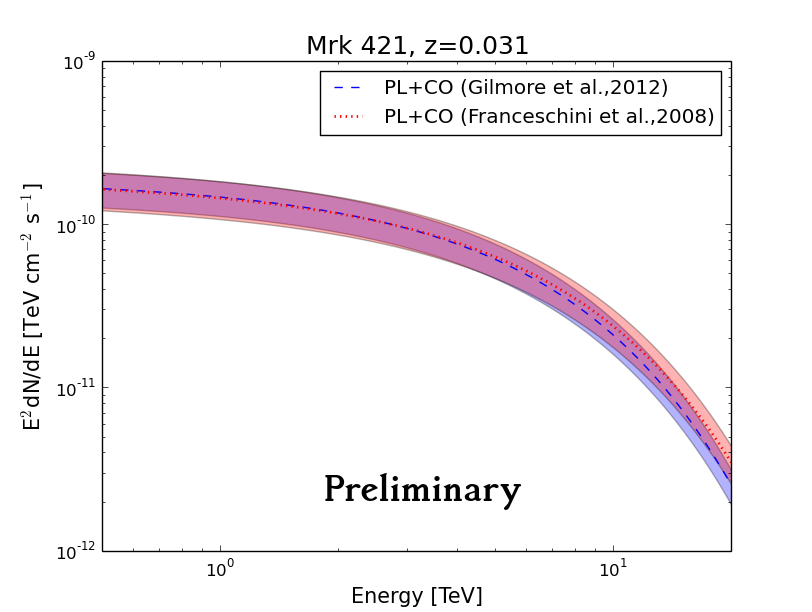}
\caption{\small{Spectral shape fit for the preferred intrinsic spectral model, PL+CO, using EBL models from \cite{Gil} (blue line) and \cite{Fran} (red line). The shadowed area is the statistical uncertainty range.}}\label{421plot}
\end{figure}

The values in Table \ref{mrk421_global} are compatible with the ones presented in \cite{FACTHAWC}. Since the $\Delta\mbox{TS}$ values are significant we can claim that the energy cut-off in the observations are not only due to the EBL and there is an intrinsic attenuation effect in Mrk 421. The Figure \ref{421plot} shows the fit of the intrinsic spectrum, a PL+CO, using EBL models from \cite{Gil} (blue line) and \cite{Fran}  (red line). The shadowed area is the statistical uncertainty range.

\subsection{Mrk 501}
We used the EBL models from \cite{Gil} and \cite{Fran} to fit the spectral shape assuming two different spectral shapes: a PL and a PL+CO. For both EBL models there is no a significant improvement with and without an exponential energy cut-off ($\Delta\mbox{TS}<1$). The result of the fits is shown in Table \ref{mrk501_global}. 

\begin{table}[h]
\begin{center}
\begin{tabular}{ccc}
\hline 
EBL model & $\Gamma$ & $\Delta\mbox{TS}$\\ 
\hline 
Gilmore et al. (2012)     & $2.25\pm0.04$ & 0.04\\
Franceschini et al. (2008)& $2.15\pm0.06$ & 0.21\\
\hline 
\end{tabular} 
\end{center}
\caption{\small{Fit results using all the energy bins for Mrk 501 with EBL models from Gilmore et al. 2012 \cite{Gil} and Franceschini et al. 2008 \cite{Fran}, $\Gamma$ is the spectral index and $\Delta\mbox{TS}$ is the TS difference between the spectral models assumptions, in this case between a PL and a PL+CO.}}\label{mrk501_global}
\end{table}

\begin{figure}
\centering
\includegraphics[scale=0.5]{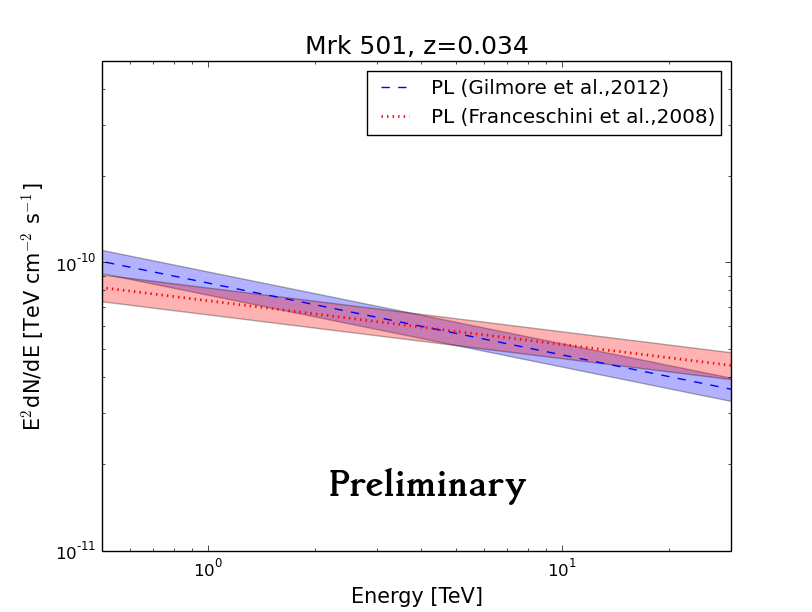}
\caption{\small{Spectral shape fit for the preferred intrinsic spectral model, a PL, using EBL models from \cite{Gil} (blue line) and \cite{Fran} (red line). The shadowed area is the statistical uncertainty range.}}\label{501plot}
\end{figure}

The values in Table \ref{mrk501_global} are compatible in the boundaries with the ones presented in \cite{FACTHAWC}. Since $\Delta\mbox{TS}<1$ (for both EBL models) we can assume that there is no intrinsic curvature in Mrk 501 and the only attenuation effect is due to the EBL. The Figure \ref{501plot} shows fit for a PL spectral model, using EBL models from \cite{Gil} (blue line) and \cite{Fran} (red line). The shadowed area is the statistical uncertainty range.

\section{Summary and Future work}\label{Conclusions}

Markarian 421 and Markarian 501 are the nearest HSP BL Lacs at $z=0.031$ and $z=0.034$ respectively and have been detected by HAWC. We present the spectral analysis of these sources using a dataset that goes from November 2014 to June 2016 comprising 543 transits. The spectral analysis was performed using an energy estimator based on an artificial neural network employed to divide the data into energy bins \cite{NN}, and the HAWC maximum-likelihood framework \cite{LiFF}.

Using EBL models from \cite{Gil} and \cite{Fran}, we show preliminary results. For Mrk 421 the spectra has a curved shape which may indicate that there is some intrinsic attenuation process, while for Mrk 501 a simple power law can well describe its intrinsic spectra. These preliminary results are in agreement with the ones presented in \cite{FACTHAWC} for Mrk 421 and in the boundaries for Mrk 501. 

Since both sources are the same type of AGN  and are practically at the same distance, the EBL effect is similar and the fact that we are showing predicted different intrinsic spectra give us the starting point to study in more detail the physics process in the emission mechanisms in BL Lacs. 

Since the NN-energy estimator is still in development, the systematics study must be the next step considering a wider range of EBL models in order to be tested and constrain the EBL intensity in the $\lambda\approx10-70$ $\mu$m wavelength region, in addition with a bigger dataset. And as mention in \cite{LC}, Mrk 501 has had various activity states during the first 17 months of observations, so one of the next objectives is to characterize the spectra of these sources during flares.

\section{Acknowledgements}
We acknowledge the support from: the US National Science Foundation (NSF); the US Department of Energy Office of High-Energy Physics; the Laboratory Directed Research and Development (LDRD) program of Los Alamos National Laboratory; Consejo Nacional de Ciencia y Tecnolog\'{\i}a (CONACyT), M{\'e}xico (grants 271051, 232656, 260378, 179588, 239762, 254964, 271737, 258865, 243290, 132197), Laboratorio Nacional HAWC de rayos gamma; L'OREAL Fellowship for Women in Science 2014; Red HAWC, M{\'e}xico; DGAPA-UNAM (grants IG100317, IN111315, IN111716-3, IA102715, 109916, IA102917); VIEP-BUAP; PIFI 2012, 2013, PROFOCIE 2014, 2015;the University of Wisconsin Alumni Research Foundation; the Institute of Geophysics, Planetary Physics, and Signatures at Los Alamos National Laboratory; Polish Science Centre grant DEC-2014/13/B/ST9/945; Coordinaci{\'o}n de la Investigaci{\'o}n Cient\'{\i}fica de la Universidad Michoacana. Thanks to Luciano D\'{\i}az and Eduardo Murrieta for technical support.

\bibliographystyle{JHEP}
\bibliography{biblio}



\end{document}